\documentclass[reprint,twocolumn,floats,english,prb,aps,preprintnumbers,amsmath,amssymb,array]{revtex4-1}
\usepackage[T1]{fontenc}
\usepackage[latin9]{inputenc}\usepackage{verbatim}
\usepackage{float}
\usepackage{graphicx}

\makeatletter
\@ifundefined{textcolor}{}
{%
 \definecolor{BLACK}{gray}{0}
 \definecolor{WHITE}{gray}{1}
 \definecolor{RED}{rgb}{1,0,0}
 \definecolor{GREEN}{rgb}{0,1,0}
 \definecolor{BLUE}{rgb}{0,0,1}
 \definecolor{CYAN}{cmyk}{1,0,0,0}
 \definecolor{MAGENTA}{cmyk}{0,1,0,0}
 \definecolor{YELLOW}{cmyk}{0,0,1,0}
 }

\usepackage{dcolumn}% Align table columns on decimal point
\usepackage{bm}% bold math
\usepackage{enumerate}

\makeatother

\usepackage{babel}

\begin{document}

\title{Analytical Continuation Approaches to Electronic Transport: The Resonant
Level Model}

\author{Eli Y. Wilner}
\affiliation{School of Physics and Astronomy, The Sackler Faculty of Exact Sciences,
Tel Aviv University,Tel Aviv 69978,Israel }

\author{Tal J. Levy}
\affiliation{School of Chemistry, The Sackler Faculty of Exact Sciences, Tel Aviv
University,Tel Aviv 69978,Israel}

\author{Eran Rabani}
\affiliation{School of Chemistry, The Sackler Faculty of Exact Sciences, Tel Aviv
University,Tel Aviv 69978,Israel}

\date{\today}

\begin{abstract}
The analytical continuation average spectrum method (ASM) and maximum
entropy (MaxEnt) method are applied to the dynamic response of a
noninteracting resonant level model within the framework of the Kubo
formula for electric conductivity. The frequency dependent
conductivity is inferred from the imaginary time current-current
correlation function for a wide range of temperatures, gate voltages
and spectral densities representing the leads, and compared with exact
results. We find that the MaxEnt provides more accurate results
compared to the ASM over the full spectral range.
\end{abstract}
\maketitle

\section{Introduction}
\label{sec:intro}
The computation of real time correlation functions in many-body
quantum systems is challenging due to the exponential complexity of
evaluating exact quantum
dynamics.\cite{loh_sign_1990,yanagisawa_quantum_2007,bouadim_sign_2008,during_statistical_2010}
This is exemplified by the well-known dynamical sign problem common to
real-time Monte-Carlo techniques. The sign problem can be avoided in
imaginary time and Wick rotation may be used to recover all the real
time information and excitation spectra. However, since the imaginary
time correlation function must be determined numerically via Quantum
Monte Carlo (QMC),\cite{Berne86rb} the rotation becomes numerically
unstable and is highly sensitive to statistical
errors.\cite{jarrell_bayesian_1996}

Several approaches have been developed in order to circumvent this
problem, such as the maximum entropy (MaxEnt) method where the optimal
fitting of a data is defined in a Bayesian manner, which in general
describes the competition between the $\chi^{2}$ goodness of the fit
and an entropic prior $S$.\cite{jarrell_bayesian_1996,Krilov01c} The
MaxEnt approach has been applied successfully to a number of
physically interesting
problems.\cite{Gubernatis90,gubernatis_quantum_1991,gubernatis_quantum_1991-1,Berne94a,Ceperley96,Berne96a,Krilov99,Rabani00,Krilov01a,Krilov01b,Rabani02c,Rabani03,Rabani05,Manolopoulos07,Miller08,Voth2008}
A second approach is based on the notion of averaging over a sequences
of possible solutions. Such approach is called stochastic analytical
continuation method or average spectrum methods
(ASM).\cite{sandvik_stochastic_1998,assaad_spin_2008,syljuasen_using_2008,kletenik-edelman_analytic_2010}
Recent application of the ASM and MaxEnt approaches argued that the
ASM should be superior to the MaxEnt at least in its ability to
resolve sharp spectral
features.\cite{vitali_path-integral_2008,reichman_analytic_2009}
Examples included the calculation of the dynamical density
fluctuations in liquid \emph{para}-hydrogen and
\emph{ortho}-deuterium~\cite{reichman_analytic_2009} as well as spin
dynamics in anti ferromagnetic Heisenberg spin
chain.\cite{syljuasen_using_2008}

In this work we apply both methods to study the dynamic response of
the well known resonant level
model~\cite{stovneng_buttiker-landauer_1989} and exhibit the dc
conductivity extracted from these approaches as well its frequency
dependence.  Our approach to describe the conductivity is quite
different from that discussed recently in the
literature,\cite{Han2010} in that it is not limited to a specific
choice of the Hamiltonian; it does however, apply to equilibrium
situations only.  To assess the accuracy of both analytic continuation
methods, we compare the results to the exact solution, which predicts
a smooth broad spectrum. We find the MaxEnt approach provides an
overall good agreement with the exact solution for the entire range of
frequencies and is particularly accurate at low frequencies near the
dc regime. On the other hand, the ASM yields sharp, narrowed spectral
features as well as spurious concave domains which are noticeably
different from the exact results. Moreover, its low frequency
predictions are somewhat less accurate compared to those of the MaxEnt
method.

The paper is organized as follows: In Section~\ref{sec:maxent} we
describe an analytic continuation approach for electrical conductivity
and briefly review two techniques to perform the Wick rotation: The
ASM and the MaxEnt method.  In Section~\ref{sec:model} we describe our
model Hamiltonian and the exact solution for the electrical
conductivity within Kubo's linear response theory.
Section~\ref{sec:results} is devoted to present the results obtained
from the ASM and MaxEnt approaches. Conclusions are given in
Section~\ref{sec:conc}.

\section{Analytical Continuation}
\label{sec:maxent}
The objective is to obtain the electrical conductivity from Kubo's
linear response theory:\cite{kubo_statistical-mechanical_1957}
\begin{equation}
\sigma_{\kappa}(\omega) = \beta \int_{0}^{\infty} dt e^{-i \omega t} \langle
\hat{I}(t) \hat{I}(0) \rangle_{\kappa}
\label{eq:kubo-sigma}
\end{equation}
where $\beta=\frac{1}{k_{\mbox{\tiny B}}T}$ is the inverse
temperature, $\hat{I}$ is the current operator defined below (see
Eq.~\ref{eq:current}), and the label $\langle \cdots \rangle_{\kappa}$
represents the Kubo transform of the current-current correlation
function defined by:\cite{kubo_statistical-mechanical_1957}
\begin{equation}
\langle \hat{I}(t) \hat{I}(0) \rangle_{\kappa} \equiv
\frac{1}{\beta\hbar} \int_{0}^{\beta\hbar} d\lambda \langle
\hat{I}(t-i\lambda) \hat{I}(0) \rangle.
\label{eq:kubo-transform}
\end{equation}
In the above, $\hbar$ is the reduced Planck's constant and $\langle
\cdots \rangle$ denotes an ensemble average.

To obtain the electrical conductivity, one requires the calculation of
the current-current correlation function, $C\left(t\right)=\langle
\hat{I}(t) \hat{I}(0) \rangle$, in real time:
\begin{equation}
C\left(t\right)=\frac{1}{Q}Tr\left\{ e^{-\beta\hat{H}}
e^{\frac{it}{\hbar}\hat{H}} \hat{I}
e^{-\frac{it}{\hbar}\hat{H}}\hat{I}\right\},
\label{eq:auto corr}
\end{equation}
where $\hat{H}$ is the Hamiltonian of the problem and $Q$ the canonical
partition function. $C\left(t\right)$ is related to the frequency
response function $D\left(\omega\right)$ by a simple Fourier relation:
\begin{equation}
C\left(t\right)=\frac{1}{2\pi}\overset{\infty}{\underset{-\infty}{\int}}e^{-i\omega
  t}D\left(\omega\right)d\omega.
\label{eq:autocorr_fourier_transform_to_time_domain}
\end{equation}
The frequency response function, $D\left(\omega\right)$, can be
written in terms of the real part of the electrical conductivity,
$\sigma_{\kappa}(\omega)$:
\begin{equation}
D\left(\omega\right) = \hbar\omega
\left(1+\coth\left(\frac{\beta\hbar\omega}{2}\right)\right)
\Re\{\sigma_{\kappa}\left(\omega\right)\},
\label{eq:realtion_structer}
\end{equation}
where for $\omega\rightarrow0$, the dc conductivity is given by
($\Im\{\sigma_{\kappa}\left(\omega \rightarrow 0\right)\}=0$):
\begin{equation}
\sigma_{\kappa}\left(0\right) = \frac{\beta}{2}D\left(0\right).
\label{eq:dc}
\end{equation}

To obtain $\sigma\left(\omega\right)$, one requires the direct
calculation of $C\left(t\right)$, which is an extremely difficult task
for a typical model Hamiltonian used to describe transport in confined
systems.  However, in comparison the real time case, the calculate the
corresponding imaginary-time function by QMC techniques is less
tedious.\cite{Ceperley95} In imaginary time, the corresponding
autocorrelation function is also related to
$D\left(\omega\right)$. This relation is achieved by performing the
Wick rotation, replacing $t\rightarrow-i\tau$ and using the detailed
balance relation
$D\left(-\omega\right)=e^{-\beta\hbar\omega}D\left(\omega\right)$:
\begin{equation}
C\left(-i\tau\right)=\frac{1}{2\pi}\underset{0}{\overset{\infty}{\int}}\left(e^{\left(\tau-\beta\hbar\right)\omega}+e^{-\omega\tau}\right)D\left(\omega\right)d\omega,
\label{eq:detail_balance_realtion}
\end{equation}
where $0\leq\tau\leq\hbar\beta$ and the imaginary time autocorrelation
function can be calculated by modifying the Heisenberg equation of
motion in Eq.~(\ref{eq:auto corr}):
\begin{equation}
C\left(-i\tau\right) = \frac{1}{Q}
Tr\left\{e^{-\beta\hat{H}}e^{\frac{\tau}{\hbar}\hat{H}}\hat{I}
e^{-\frac{\tau}{\hbar}\hat{H}}\hat{I}\right\}.
\label{eq:imaginary_time_auto_corr_function}
\end{equation}

Much of the difficulty in calculating $C\left(t\right)$ is now shifted
to that of inverting the above relation to obtain $D(\omega)$ from
$C(-i\tau)$.  We refer to two approaches developed to invert this
relation: The MaxEnt method and the ASM. The MaxEnt method selects the
solution which maximizes the posterior probability, or the probability
of the solution $D(\omega)$ given a data set $C(-i\tau)$.  Using
Bayes' theorem, one can show that the posterior probability is given
by~\cite{Gubernatis91,Skilling89}
\begin{equation}
{\cal{P}}(D(\omega)|C(-i\tau)) \propto \exp(\alpha S - \chi^2/2).
\label{eq:posteriorprob}
\end{equation}
where $S$ is the information entropy and $\chi^2$ is the standard mean
squared deviation from the data.  In the present study, we only
consider the standard L-curve method to determine
$\alpha$.\cite{Lawson95} In this context we regard $\alpha$ as a
regularization parameter controlling the degree of smoothness of the
solution, and the entropy as the regularizing function. Its value is
selected by constructing a plot of $\log[-S]$ vs. $\log \chi^2$. This
curve has a characteristic L-shape, and the corner of the L, or the
point of maximum curvature, corresponds to the value of $\alpha$ which
is the best compromise between fitting the data and obtaining a smooth
solution. Our experience with analytic continuation has lead us to the
conclusion that this approach is the most robust.  Once the
regularization parameter is determined, we apply the approach
described in Ref.~\onlinecite{Krilov01c} to obtain the statistically
rigorous and unique fit for $D(\omega)$.

The basic idea behind the ASM is to pick the final solution for
$D(\omega)$ as the average spectral function obtained by averaging
over a posterior probability, ${\cal{P}}(D(\omega)| C(-i\tau))$,
instead of taking the value that maximizes this distribution.  Thus,
fluctuations of the solution are allowed in the ASM:
\begin{equation}
\bar{D}(\omega)=\frac{\int d|D(\omega)| D(\omega) {\cal{P}}(D(\omega)|
  C(-i\tau))} {\int d|D(\omega)| {\cal{P}}(D(\omega)|C(-i\tau))}.
\label{eq:Dwav}
\end{equation}
The averaging is performed by a Monte Carlo procedure preserving known
sum rules.\cite{reichman_analytic_2009} Readers who are interested in
a more comprehensive discussion of the ASM are referred to
Refs.~\onlinecite{Sandvik98} and \onlinecite{Syljuasen08} and to the
specific implementation of Ref.~\onlinecite{reichman_analytic_2009}.

\section{Model}
\label{sec:model}
We consider the resonant Level model, which consists of a single
quantum dot state coupled to two leads (fermionic baths). This model
has been used extensively in understanding transport properties of
non-interacting systems, most recently in a semiclassical study of
transport developed by Swenson {\em et
  al.}\cite{swenson_application_2011} Since an exact solution of the
frequency dependent conductivity is available, this model is ideal for
assessing the accuracy of the proposed analytic continuation approach.
The Hamiltonian is given by (more details can be found in
Ref.~\onlinecite{swenson_application_2011})
\begin{equation}
  \hat{H}=\varepsilon_{d}d^{\dagger}d+\underset{k \in
    L,R}{\sum}\varepsilon_{k}c_{k}^{\dagger}c_{k}+\underset{k \in
    L,R}{\sum}t_{k}\left(d^{\dagger}c_{k}+h.c.\right),
\label{eq:hamil}
\end{equation}
where $d$ ($d^{\dagger}$) is the destruction (creation) operator of an
electron on the dot, $\varepsilon_{d}$ is the energy of the isolated
dot (which will also be referred to as the gate potential), $c_k$
($c_k^{\dagger}$) is the destruction (creation) operator of an
electron on the left (L) or right (R) lead with energy
$\varepsilon_k$, and $t_{k}$ is the coupling amplitude between the dot
and the lead level $k$.

For a full specification of the model, one requires the spectral
function which determines $t_k$. In the applications reported below,
the leads will be described within the wide band limit with a sharp
cutoff at high and low energy values, such that the spectral density
is given by
\begin{equation}
  J_{L/R}\left(\varepsilon\right)=\frac{\Gamma_{L/R}}
  {\left(1+e^{A\left(\varepsilon-B\right)}\right)\left(1+e^{-A\left(\varepsilon+B\right)}\right)}
\label{eq:widebandlimit}
\end{equation}
In the calculations reported below, unless otherwise noted, we use
$\Gamma_{L}=\Gamma_{R}=\frac{1}{2}$, $\Gamma=\Gamma_{L}+\Gamma_{L}$,
$A=5\Gamma$,$B= 3-10\Gamma$.  We also use a uniform discretization
($\delta\varepsilon_{K}$) to select the energies of the leads' states,
and thus the couplings are given by:
\begin{equation}
t_{k}\left(\varepsilon_{k}\right)=\sqrt{\frac{J\left(\varepsilon_{k}\right)\Delta\varepsilon_{k}}{2\pi}}.
\label{eq:t_coupling}
\end{equation}

Our aim is to use the ASM and MaxEnt method to calculate the
measurable dc conductivity and its frequency dependence and compare
the analytic continuation approaches to exact results.  For the model
Hamiltonian given above, the current operator, $\hat{I}$, is taken to
represent the current from the left lead to the dot (similarly, one
can look at the right current), and is given by the change in
occupancy of the left lead, $\hat{I}=e\frac{d\hat{N}_{L}}{dt}$, where
$\hat{N}_{L}=\underset{k \in L}\sum c_{k}^{\dagger} c_{k}$, and a
brief calculations shows that
\begin{equation}
  \hat{I}=\frac{ie}{\hbar}\underset{k \in
    L}{\sum}t_{k}\left(c_{k}^{\dagger}d-d^{\dagger}c_{k}\right).
\label{eq:current}
\end{equation}

With the above definition, the current-current correlation function
can be calculated exactly in order to obtain the conductivity based on
Kubo's linear response theory.\cite{kubo_statistical-mechanical_1957}
Let us assume that one knows the matrix which diagonalizes the
Hamiltonian in Eq.~(\ref{eq:hamil}) and denote it by $U$. By making a
linear transformation for the operators
$c_{k}=\underset{\alpha}{\sum}U_{k\alpha}\tilde{c}_{\alpha}$ and
$c_{k}^{\dagger}=\underset{\alpha}{\sum}U_{\alpha
  k}^{-1}\tilde{c}_{\alpha}^{\dagger}$ (where from now on the index
$k=0$ will be refereed to that of the dot level), we can rewrite the
expression for $C\left(t\right)$ in terms of the eigenfunctions of
diagonal Hamiltonian:
\begin{eqnarray}
C\left(t\right) & = & \frac{1}{Q}Tr\left\{
e^{-\beta\hat{H}}e^{i\hat{H}\frac{t}{\hbar}}\hat{I}e^{-i\hat{H}\frac{t}{\hbar}}\hat{I}\right\}
\nonumber \\ & = & \frac{e^{2}}{\hbar^{2}}\underset{nm}{\sum}
e^{i\left(\tilde{\varepsilon}_{n}-\tilde{\varepsilon}_{m}\right)\frac{t}{\hbar}}
\Xi{}_{nm}^{2}f\left(\tilde{\varepsilon}_{n}\right)
\left[1-f\left(\tilde{\varepsilon}_{m}\right)\right].
\label{eq:current_current_correlation function}
\end{eqnarray}
In the above, $\tilde{\varepsilon}_{n}$ are the energy eigenvalues
from the diagonalized Hamiltonian $\tilde{H} =
\overset{\infty}{\underset{i=0}{\sum}} \tilde{\varepsilon}_{i}
\tilde{c}_{i}^{\dagger}\tilde{c}_{i}$,
$f\left(\tilde{\varepsilon}_{n}\right)$ are the Fermi-Dirac
distribution at equilibrium, and
\begin{equation}
\Xi{}_{nm}^{2} =
\underset{q,k}{\sum}t_{k}t_{q}
\left(U_{m0}^{-1}U_{kn}-U_{mk}^{-1}U_{0n}\right)
\left(U_{m0}^{-1}U_{qn}-U_{mq}^{-1}U_{0n}\right)
\end{equation}
can be refereed to as the electron velocity matrix element in the the
orthogonal basis, which obeys $\Xi_{nn}=0$ for the same quantum
state. The Fourier transform of
Eq.~(\ref{eq:current_current_correlation function}) gives
\begin{eqnarray}
D\left(\omega\right) &=& \underset{\eta\rightarrow0^{+}}\lim \frac{2\pi
  e^{2}}{\hbar} \underset{nm} {\sum} \Xi_{nm}^{2}
\frac{e^{\beta\tilde{\varepsilon}_{n}}f
  \left(\tilde{\varepsilon}_{n}\right)}
     {e^{\beta\left(\tilde{\varepsilon}_{n}-\hbar\omega\right)}+1}
     \delta_{\eta}(\Delta_{nm}+\hbar\omega),\nonumber \\ &&
\label{eq:omega_space_correlation_function}
\end{eqnarray}
where $\Delta_{nm}=\tilde{\varepsilon}_{n}-\tilde{\varepsilon}_{m}$
and $\delta_{\eta}(x)$ is defined by
\begin{equation}
\delta_{\eta}(x) = \frac{1}{\pi}\frac{\hbar\eta}{x^{2} +
  \left(\hbar\eta\right)^{2}}.
\end{equation}

Using Eq.~(\ref{eq:realtion_structer}), the experimental measured
electrical conductivity can be calculated and is given by:
\begin{eqnarray}
\Re\left\{ \sigma_{\kappa}\left(\omega\right)\right\} & = &
\underset{\eta\rightarrow0^{+}}\lim \frac{e^{2}}{\hbar^{2}}
\underset{nm}{\sum} \Xi_{nm}^{2} \frac{\Delta f_{mn}}
         {\Delta_{mn}}\delta_{\eta}(\Delta_{nm}+\hbar\omega)\nonumber
         \\
\label{eq:Kubo_eta}
\end{eqnarray}
where $\Delta f_{mn}=f\left(\tilde{\varepsilon}_{m}\right) -
f\left(\tilde{\varepsilon}_{n}\right)$. In practice, we use a finite
value of $\eta$ since the leads (baths) are described by a finite set
of modes. We take $\hbar \eta$ to be larger than the spacing between
adjacent bath modes and smaller than the band width. Its actual value
is determined by starting from a large value of $\eta$ and gradually
decreasing it until convergence occurs.\cite{imry_introduction_2002}

Formally, for an infinite number of bath modes one can take the limit
$\eta \rightarrow 0^{+}$:\cite{imry_introduction_2002}
\begin{eqnarray}
\Re\left\{ \sigma_{\kappa}\left(\omega\right)\right\}  & = & \frac{\pi e^{2}}{2\hbar}\frac{\sinh\left(\frac{\beta\hbar\omega}{2}\right)}{\hbar\omega}\underset{nm}{\sum}\Xi_{nm}^{2}\nonumber \\
 &  & \,\times \mbox{sech}\left(\frac{\beta\tilde{\varepsilon}_{m}}{2}\right)\mbox{sech}\left(\frac{\beta\left(\tilde{\varepsilon}_{m}-\hbar\omega\right)}{2}\right)\nonumber \\
&  & \,\times\delta(\tilde{\varepsilon}_{n}-\tilde{\varepsilon}_{m}+\hbar\omega),
\label{eq:kubo_eta_0}
\end{eqnarray}
and the dc conductivity is then obtain by taking $\omega$ to $0$,
which gives the well known Landauer formula for
conductance~\cite{datta_electronic_1997,datta_quantum_2005}
\begin{equation}
  \sigma_{\kappa}(0) =\frac{\pi e^{2}}{4\hbar}\beta\underset{nm}{\sum}\Xi_{nm}^{2}\mbox{sech}^{2}\left(\frac{\beta\tilde{\varepsilon}_{m}}{2}\right)\delta(\tilde{\varepsilon}_{n}-\tilde{\varepsilon}_{m}).
\label{Landauer Formula}
\end{equation}

\section{Results}
\label{sec:results}
The application of the ASM and MaxEnt method requires as input the
imaginary time current-current correlation function, $C(-i\tau)$ given
by Eq.~\ref{eq:imaginary_time_auto_corr_function}. In principle, the
calculation of such correlation functions requires numerical
techniques such as those based on path integration, where a variety of
approaches can be extended to imaginary
time.\cite{hirsch_monte_1986,werner_continuous-time_2006,muehlbacher_real-time_2008,weiss_iterative_2008,werner_diagrammatic_2009,schiro_real-time_2009,segal_numerically_2010,werner_weak-coupling_2010,Gull10}
In the present case, due to the simplicity of the model Hamiltonian
(cf., Eq.~(\ref{eq:hamil})), instead of referring to a specific
numerical implementation of a path integration approach, we derive an
exact expression for $C(-i\tau)$.  To preform the analytic
continuation based on the ASM and MaxEnt method, we add artificial
noise to the exact expression of $C(-i\tau)$ to mimic the results of a
Monte Carlo procedure, as described below.

For a finite value of $\eta$ (see the discussion above), the imaginary
time current-current correlation function is given by the exact
expression:
\begin{widetext}
\begin{eqnarray}
C_{\eta}(\tau) & = &
-\frac{ie^{2}}{2\pi\hbar^{2}}\underset{nm}{\sum}\Lambda_{nm}
\left(e^{\Omega_{mn}\tau}E_{1}\left(\Omega_{mn}\tau\right)-e^{\Omega_{mn}^{*}\tau}E_{1}
\left(\Omega_{mn}^{*}\tau\right)\right.\nonumber \\ & &
\left.+e^{\Omega_{mn}^{*}(\beta\hbar-\tau)}\left\{
E_{i}\left(\Omega_{mn}^{*}(\tau-\beta\hbar)\right)+i\pi\right\}
+ie^{\Omega_{mn}(\beta\hbar-\tau)}\left\{
\pi+iE_{i}\left(\Omega_{mn}(\tau-\beta\hbar)\right)\right\} \right)
\label{eq:corr_imag_eta}
\end{eqnarray}
\end{widetext}
where $\Omega_{mn} = \frac{\tilde{\varepsilon}_{m} -
  \tilde{\varepsilon}_{n}-i\hbar\eta}{\hbar},\,\,
\Lambda_{nm}=\Xi_{nm}^{2} f({\tilde{\varepsilon}_{m}})
f({\tilde{\varepsilon}_{n}})e^{\beta\tilde{\varepsilon}_{n}}$, and
$E_{i}\left(x\right) = -\underset{-x}{\overset{\infty}{\int}}
\frac{e^{-t}}{t}dt,\,\,\, E_{n}\left(x\right) =
-\underset{1}{\overset{\infty}{\int}}\frac{e^{-xt}}{t^{n}}dt$ are the
exponential integral function and the $n^{\mbox{\tiny{th}}}$
exponential integral function, respectively. The above expression for
$C_{\eta}(\tau)$ is well defined for $0 \le \tau \le \hbar \beta$ and
reduces to
\begin{equation}
C(-i\tau) \equiv C_{\eta \rightarrow 0^{+}}\left(\tau\right) =
\frac{e^{2}}{\hbar^{2}}\underset{nm}
     {\sum}e^{\left(\tilde{\varepsilon}_{n}-\tilde{\varepsilon}_{m}\right)
       \frac{\tau}{\hbar}}\Lambda_{nm}
\label{eq:corr_imag_eta+0}
\end{equation}
in the limit of $\eta \rightarrow 0^{+}$.

\begin{figure}[H]
\includegraphics[width=8cm]{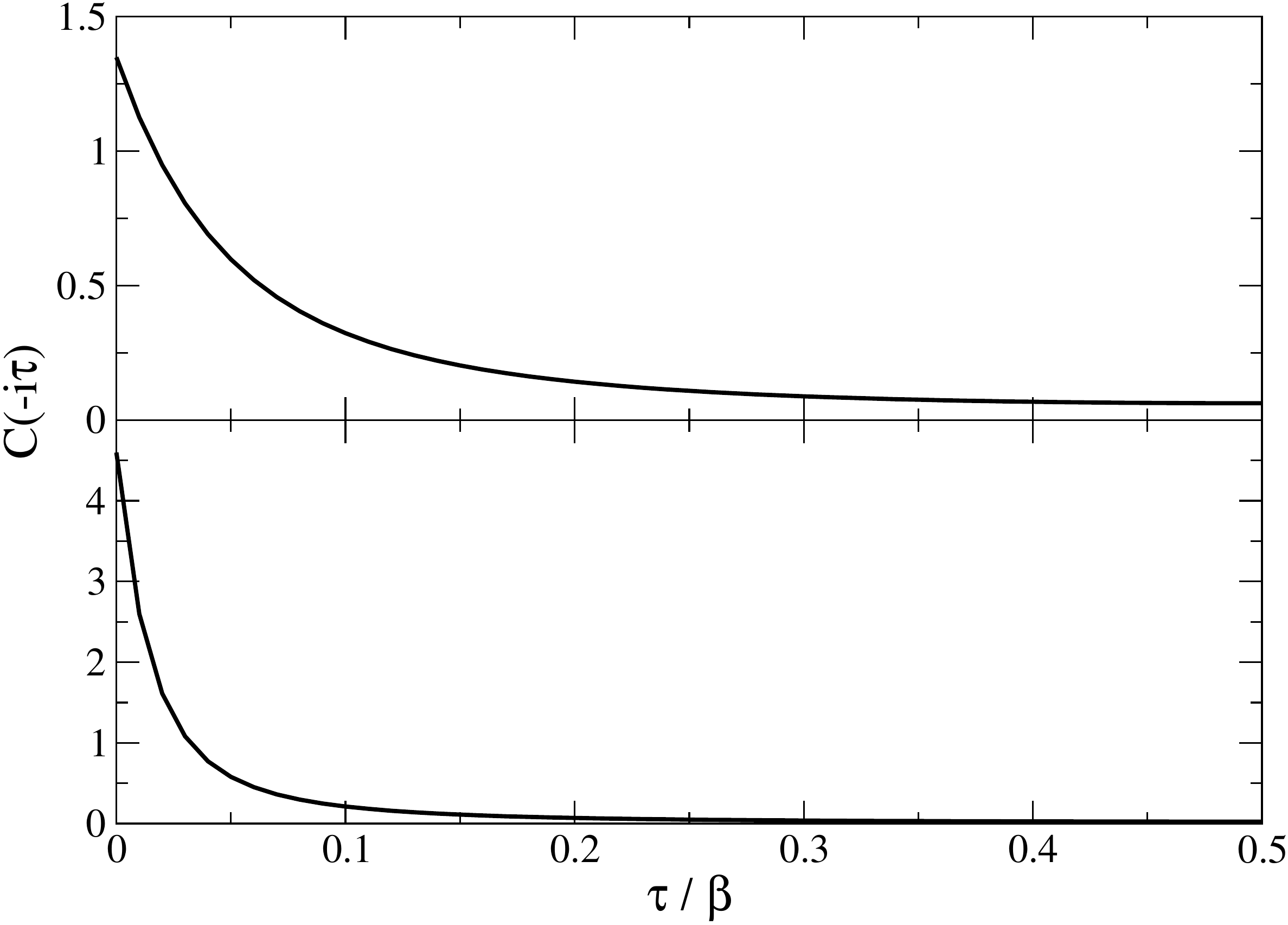}
\caption{Current-current correlation function in imaginary time (cf.,
  Eq.~(\ref{eq:corr_imag_eta}) for $\beta=3/\Gamma$ (upper panel) and
  $\beta=10/\Gamma$ (lower panel).  Results are shown for $\hbar \eta
  = 0.04 \Gamma$ and $\epsilon_{d}=0$. On the scale of the plots, it
  is difficult to differentiate between results obtained for different
  values of $\eta$.}
\label{fig:imagcorr}
\end{figure}

In Fig.~\ref{fig:imagcorr} we plot the current-current correlation
function in imaginary time for two values of $\beta$. The results were
obtained for a finite number of bath modes ($N_b$), where $N_{b}=1000$
for both the left ($L$) and right ($R$) leads.  The results are shown
for $\hbar \eta = 0.04 \Gamma$.  However, we find that within the
accuracy of the numerical analytic continuation, the inversion is not
sensitive to the value of $\eta$. In fact, on the scale of the plots
shown in Fig.~\ref{fig:imagcorr}, different values of $\eta$ (within a
reasonable range) are indistinguishable.  This implies that one can
refer to Monte Carlo simulations to obtain the imaginary time data for
a finite number of bath modes and ignore the role of $\eta$ in the
simulations.  This, of course, is not the case in real time, where for
a finite system $\sigma_{\kappa}(\omega = 0) \rightarrow 0$ when
$\eta=0$.\cite{imry_introduction_2002}

\begin{figure}[H]
\includegraphics[width=8cm]{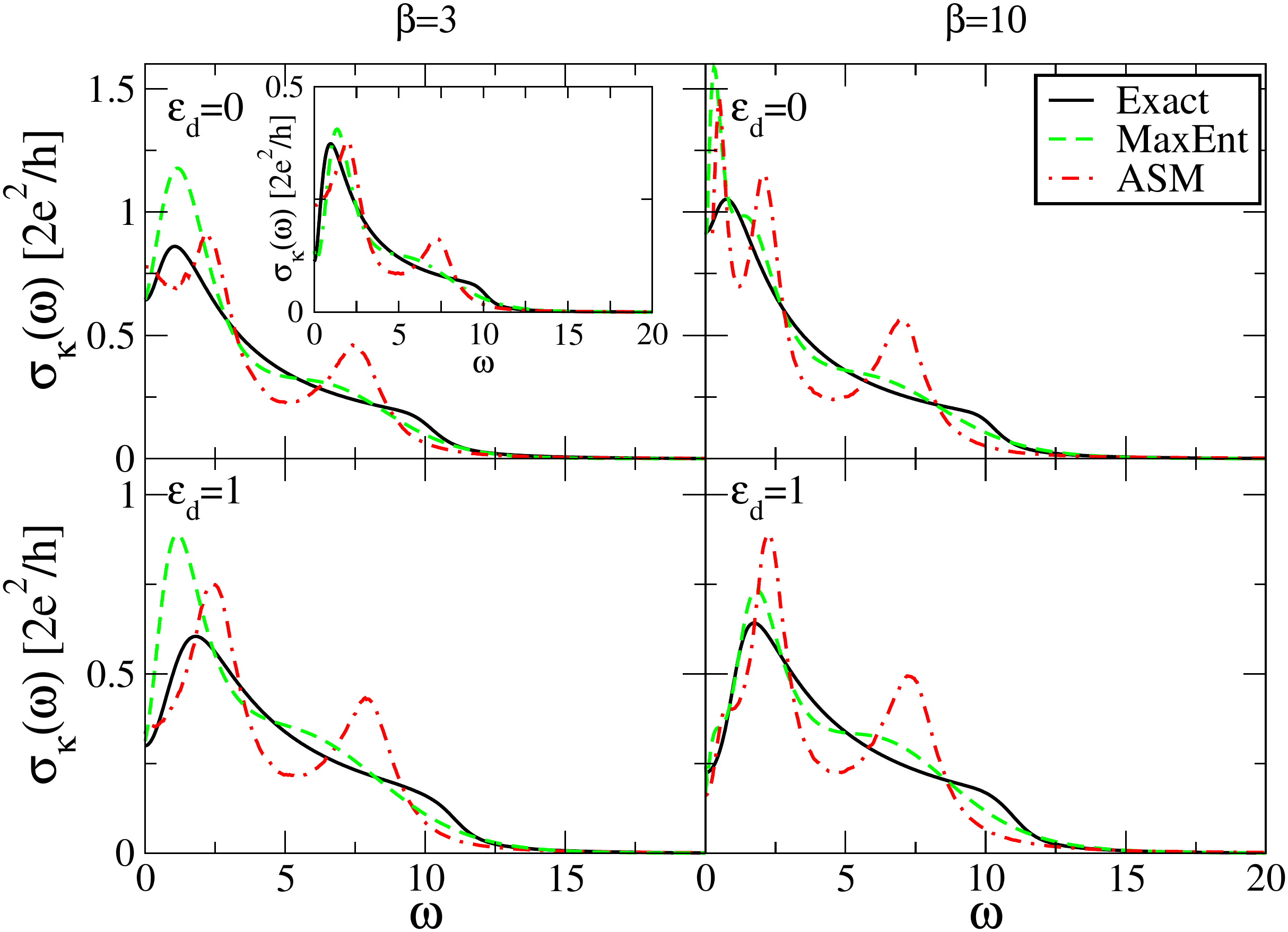}
\caption{Frequency dependent electrical conductivity
  $\sigma_{\kappa}(\omega)$ obtained from the exact result from
  Eq.~(\ref{eq:Kubo_eta}) (solid black curve), the ASM (dashed-dotted
  red curve) and the MaxEnt (dashed green curve) for different gate
  voltages $\left(\varepsilon_{d}\right)$ and
  temperatures. $\epsilon_d$ is denoted in units of $\Gamma$, $\beta$
  in units of $1/\Gamma$, and $\omega$ in units of
  $\Gamma/\hbar$. Inset: Frequency dependent electrical conductivity
  $\sigma_{\kappa}(\omega)$ for the asymmetric coupling case $\Gamma_L
  = \frac{1}{2}$ and $\Gamma_R = \frac{1}{10}$.}
\label{fig:sigmaw}
\end{figure}

Next, we applied both the ASM and MaxEnt method to invert the
imaginary time data. We added Gaussian noise to the imaginary time
current-current correlation function with a standard deviation of
$1\%$ at each data point. We discretized the imaginary time axis to
$N_{\tau}=100$ points and the frequency axis to $N_{\omega}=1024$
points.  For the ASM approach, we averaged the solution over $10^{7}$
Monte Carlo sweeps. In Fig.~\ref{fig:sigmaw} we plot the results for
the frequency dependent electrical conductivity at a gate voltage
$\varepsilon_{d}=0$ and $\varepsilon_{d}=\Gamma$, and for two inverse
temperatures of $\beta=\frac{3}{\Gamma}$ and
$\beta=\frac{10}{\Gamma}$.

We find that the MaxEnt provides an overall better agreement with the
exact results compared to the ASM. Both approaches provide reasonable
description at low frequencies, and thus are accurate enough to
determine the dc conductivity (see below), though MaxEnt is in general
more accurate in this spectral regime.  Both approaches over-estimate
the magnitude of the first peak observed at low frequencies. In
general, MaxEnt does provide a better estimate of the position, height
and width of the peak for the range of model parameters studied in
this work. The ASM also produces a sharp peak at higher frequencies
which does not appear in the exact solution for
$\sigma_{\kappa}(\omega)$, while MaxEnt performs better in this
regime. In fact, the high frequency behavior is quite tough for
analytic continuation approaches.  In $D(\omega)$ (not shown) we
observe a sharp fall of the response as the frequency approaches the
band cutoff $B$. This is translated to a small shoulder observed at
$\hbar\omega=B$ in $\sigma(\omega)$. Such a sharp change in the
spectral behavior is challenging for the analytic continuation
methods, and this explains the failure of the ASM.

The inset of Fig.~\ref{fig:sigmaw} shows a typical result for the
asymmetric coupling case, where $\Gamma_L = \frac{1}{2}$ and $\Gamma_R
= \frac{1}{10}$. This case is more difficult for analytic continuation
since the overall magnitude of the conductivity is much smaller, due
to the occurrence of destructive interference on the dot. MaxEnt
provides reasonable results at all frequencies while the ASM fails
markedly even at low frequencies. The fact that MaxEnt captures the
behavior of the conductivity for the asymmetric case is
encouraging. In fact, it suggests that the MaxEnt approach can be used
for more elaborate situations: For example, when the coupling to the
left lead is not proportional to that of the right lead and the simple
Landauer type Meir-Wingreen formula is not adequate (see Eq.~(9) in
Ref.~\onlinecite{meir_landauer_1992}).

\begin{figure}[H]
\includegraphics[width=8cm]{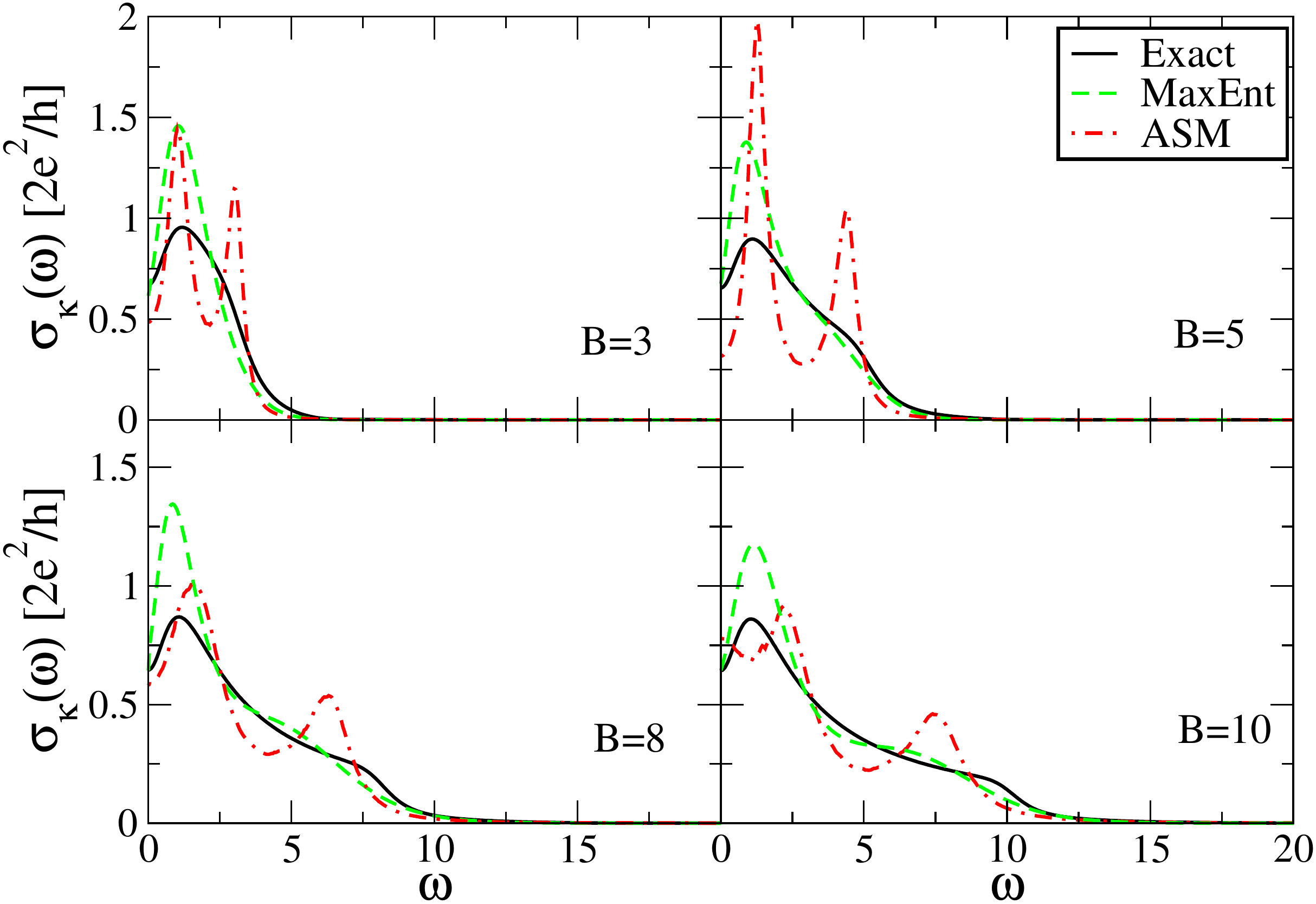}
\caption{Same as Fig.~\ref{fig:sigmaw} but for different values of the
  lead cutoff parameter, $B$, given in units of $\Gamma$.}
\label{fig:sigmawB}
\end{figure}

The high frequency behavior is also determined by the band structure
of the leads. In Fig.~\ref{fig:sigmawB} we plot the frequency
dependent electrical conductivity for different values of the band
cutoff parameters, $B$ defined in
Eq.~(\ref{eq:widebandlimit}). Indeed, the MaxEnt method gives a well
behaved spectral response while the ASM provides two peaks for all
values of $B$. This is even more pronounced at small values of $B$,
where the width of the peaks in the ASM are very narrow, in contrast
to the exact response which is smooth.  Moreover, even the low
frequency response of the ASM significantly deviates from the exact
results at low band width, while within MaxEnt the accuracy at low
frequency remains the same for all values of $B$, and the dc
conductivity is only slightly affected by changing $B$.

\begin{figure}[H]
\includegraphics[width=8cm]{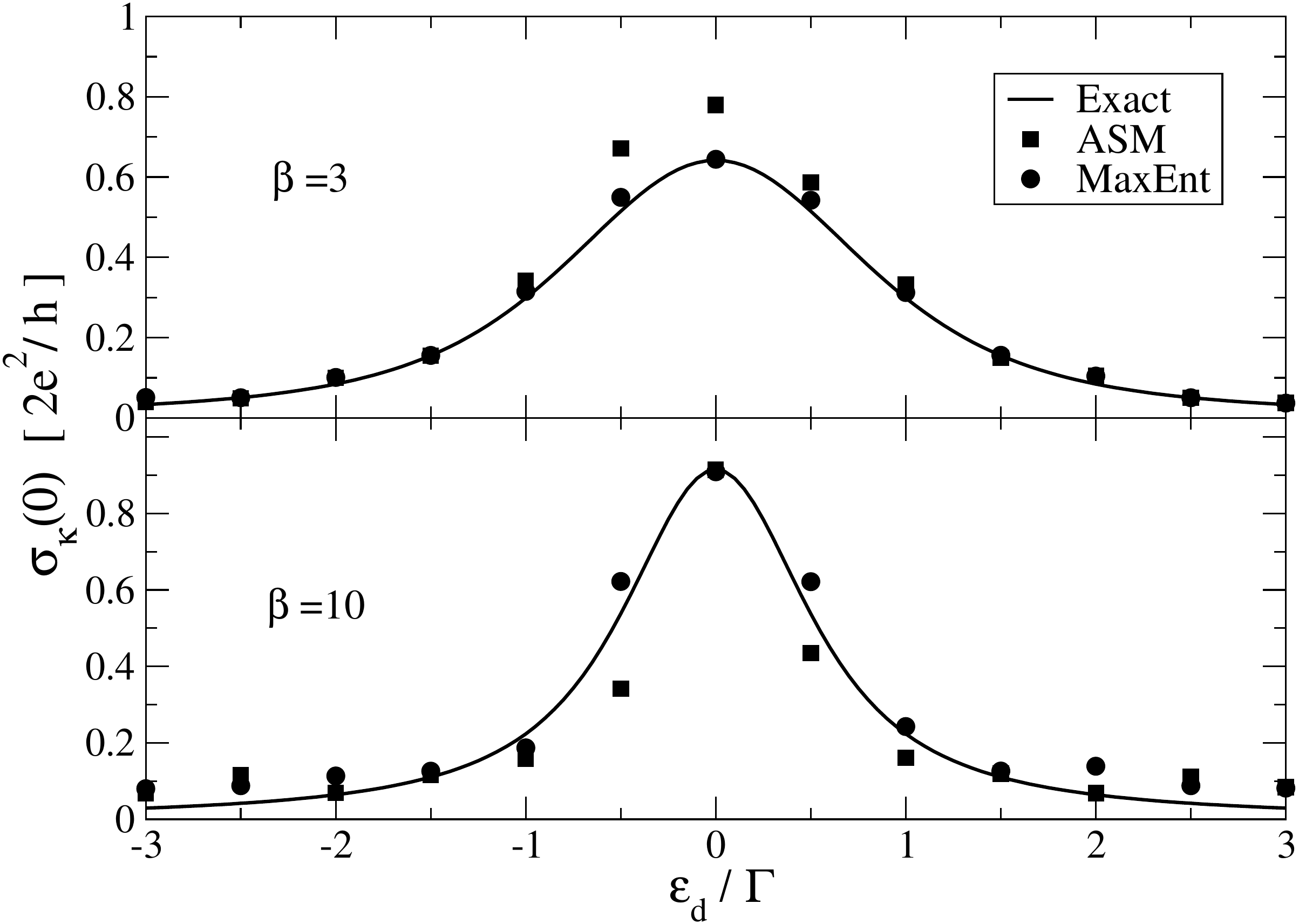}
\caption{The dc conductivity $\sigma_{\kappa}(\omega = 0)$ as a
  function of the gate voltage $\epsilon_d$ for temperature
  $\beta=3/\Gamma$ (upper panel) and $\beta=10/\Gamma$ (lower panel).}
\label{fig:dcgate}
\end{figure}

\begin{figure}[H]
\includegraphics[width=8cm]{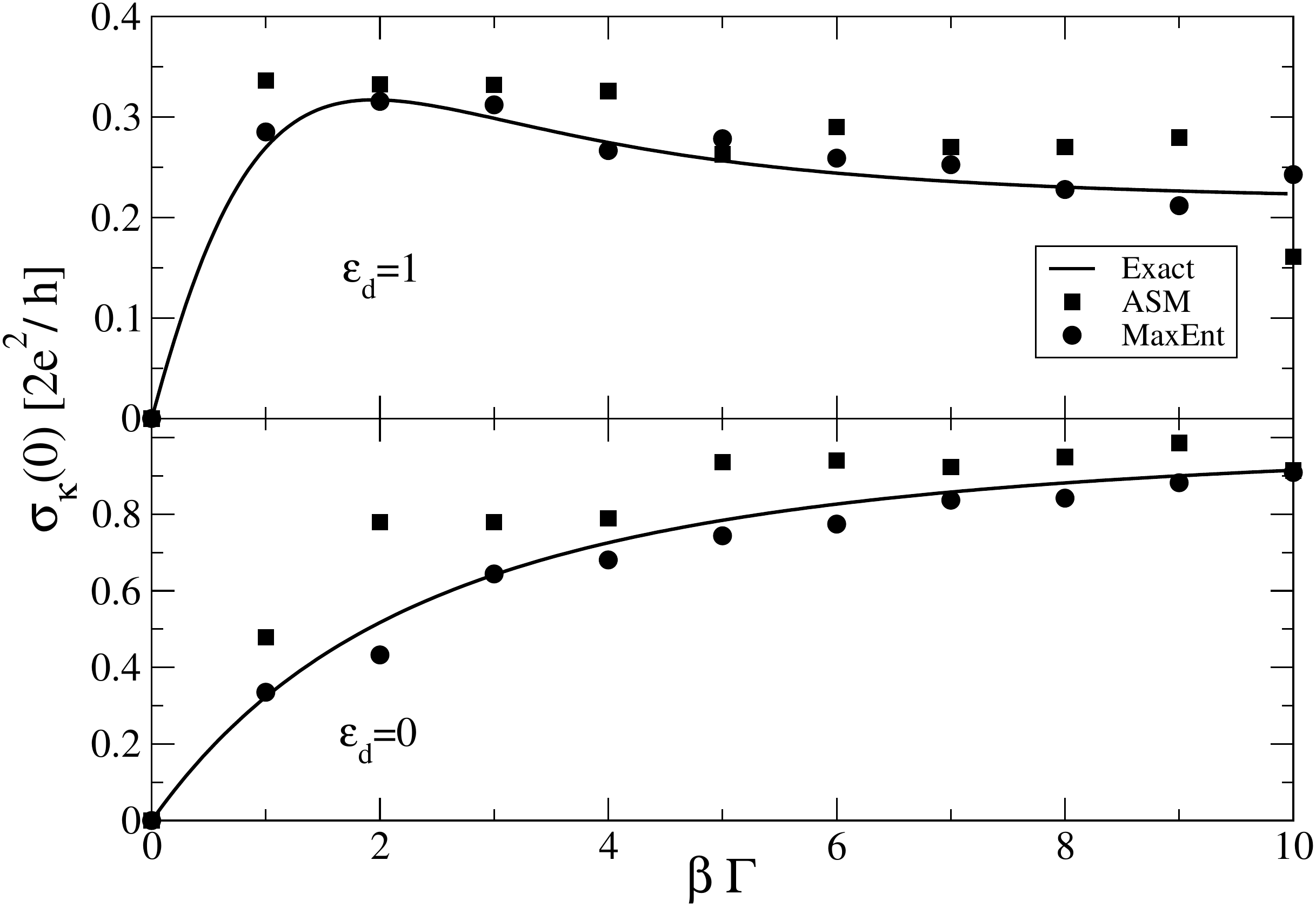}
\caption{The dc conductivity as a function of the inverse temperature
  for gate voltage $\varepsilon_{d}=\Gamma$ (upper panel) and
  $\varepsilon_{d}=0$ (lower panel).}
\label{fig:dcbeta}
\end{figure}

The dc conductivity as a function of the gate voltage and inverse
temperature is shown in Fig.~\ref{fig:dcgate} and
Fig.~\ref{fig:dcbeta}, respectively. The exact results given by
Eq.~(\ref{eq:Kubo_eta}) agree with the Landauer formula (not
shown).\cite{imry_introduction_2002} For the current set of
parameters, both approaches provide good results for the conductivity
at $\omega = 0$. Again, the MaxEnt method seems to provide more
accurate results near the resonance and is quantitative for most gate
voltages and temperatures.

\section{Conclusions}
\label{sec:conc}
We have presented an analytic continuation approach based on Kubo's
linear response theory to obtain the dc and ac components of the
electrical conductivity in molecular junctions. Kubo's formulation
requires the calculation of the current-current correlation function
in real time, which is a difficult task for a many-body open quantum
system due to the well-known dynamical sign problem. The calculation
of the corresponding imaginary time correlation, on the other hand, is
a much simpler task, amenable to Monte Carlo techniques.

Two approaches were adopted here to carry the analytic continuation of
the current-current correlation function to real time: the ASM and the
MaxEnt method. To assess the accuracy of these methods, we performed
calculations for the resonant level model at a wide range of
temperatures, gate voltages and frequencies. The numerical results
were compared with an exact expression for the electrical
conductivity, which is straightforward to obtain for this model.

We find that MaxEnt is superior to the ASM for the entire range of
frequencies and for different model parameters. It provides an
accurate description of the dc conductivity as well as a reasonable
approximation for the ac component.  Furthermore, MaxEnt captures
interference effects on the dot resulting from breaking the symmetry
in the couplings between the dot and the leads. The ASM fails in all
these respects.

\section{Acknowledgments }
We would like to thank Guy Cohen for helpful discussions and critical
comments on the manuscript. EYW would like to thank Dr. Kalman Wilner
for useful discussions. This work was supported by the US-Israel
Binational Science Foundation and by the FP7 Marie Curie IOF project
HJSC. TJL is grateful to the The Center for Nanoscience and
Nanotechnology at Tel Aviv University of a doctoral fellowship.

\end{document}